\documentclass[11pt]{article}         
\usepackage{amsfonts,amsmath,amssymb}

\begin{document}

\begin{center}
\textsf{\Large    On $osp(M|2n)$ integrable open spin chains}
\\[5mm]
\textbf{{D. Arnaudon}, 
N.~Cramp\'e, A.~Doikou, L.~Frappat, \'E.~Ragoucy}    \\[2mm]
{Laboratoire d'Annecy-le-Vieux de Physique Th{\'e}orique LAPTH
CNRS, UMR 5108, associ{\'e}e {\`a} l'Universit{\'e} de Savoie
LAPP, BP 110, F-74941 Annecy-le-Vieux Cedex, France} \\[2mm]
\textbf {J. Avan}   \\[2mm]
{
Laboratoire de Physique Th{\'e}orique et Mod{\'e}lisation
Universit{\'e} de Cergy, \\ 5 mail Gay-Lussac, Neuville-sur-Oise
F-95031 Cergy-Pontoise Cedex
}
\end{center}
{Pacs : 02.20.Uw, 03.65.Fd, 75.10.Pq}  

\begin{flushright}
  \textbf{LAPTH-Conf-1054/04}
\end{flushright}
\begin{abstract}
  We consider open spin chains based on $osp(m|2n)$ Yangians. We solve
  the reflection equations for some classes of reflection matrices,
  including the diagonal ones. Having then integrable open spin chains, we
  write the analytical Bethe Ansatz equations.
  More details and references can be found in \cite{yabon,banania}.
\end{abstract}

\newtheorem{definition}{Definition}[section]
\newtheorem{example}[definition]{Example}
\newtheorem{conjecture}[definition]{Conjecture}
\newtheorem{proposition}[definition]{Proposition}
\newtheorem{theorem}[definition]{Theorem}
\newtheorem{corollary}[definition]{Corollary}
\newtheorem{lemma}[definition]{Lemma}

\newcommand\textcolor[1]{{}}
\newcommand\noir[1]{\textcolor{black}{#1}}
\newcommand\rouge[1]{\textcolor{red}{#1}}
\newcommand\bleu[1]{\textcolor{blue}{#1}}
\newcommand\green[1]{\textcolor{green}{#1}}
\newcommand\dcyan[1]{\textcolor{dcyan}{#1}}
\newcommand\ddcyan[1]{\textcolor{ddcyan}{#1}}
\newcommand\jaune[1]{\textcolor{yellow}{#1}}
\newcommand\rose[1]{\textcolor{rose}{#1}}
\newcommand\magenta[1]{\textcolor{magenta}{#1}}
\newcommand\violet[1]{\textcolor{violet}{#1}}
\newcommand\dgreen[1]{\textcolor{dgreen}{#1}}

\newcommand{\II}{{\mathbb I}}
\def\CC{{\mathbb C}}
\def\NN{{\mathbb N}}
\def\QQ{{\mathbb Q}}
\def\RR{{\mathbb R}}
\def\ZZ{{\mathbb Z}}
\def\cA{{\cal A}}          \def\cB{{\cal B}}          \def\cC{{\cal C}}
\def\cD{{\cal D}}          \def\cE{{\cal E}}          \def\cF{{\cal F}}
\def\cG{{\cal G}}          \def\cH{{\cal H}}          \def\cI{{\cal I}}
\def\cJ{{\cal J}}          \def\cK{{\cal K}}          \def\cL{{\cal L}} 
\def\cM{{\cal M}}          \def\cN{{\cal N}}          \def\cO{{\cal O}}
\def\cP{{\cal P}}          \def\cQ{{\cal Q}}          \def\cR{{\cal R}} 
\def\cS{{\cal S}}          \def\cT{{\cal T}}          \def\cU{{\cal U}}
\def\cV{{\cal V}}          \def\cW{{\cal W}}          \def\cX{{\cal X}}
\def\cY{{\cal Y}}          \def\cZ{{\cal Z}}
\def\bA{{\bar A}}          \def\bB{{\bar B}}          \def\bC{{\bar C}}
\def\bD{{\bar D}}          \def\bE{{\bar E}}          \def\bF{{\bar F}}
\def\bG{{\bar G}}          \def\bH{{\bar H}}          \def\bI{{\bar I}}
\def\bJ{{\bar J}}          \def\bK{{\bar K}}          \def\bL{{\bar L}} 
\def\bM{{\bar M}}          \def\bN{{\bar N}}          \def\bO{{\bar O}}
\def\bP{{\bar P}}          \def\bQ{{\bar Q}}          \def\bR{{\bar R}} 
\def\bS{{\bar S}}          \def\bT{{\bar T}}          \def\bU{{\bar U}}
\def\bV{{\bar V}}          \def\bW{{\bar W}}          \def\bX{{\bar X}}
\def\bY{{\bar Y}}          \def\bZ{{\bar Z}}
\newcommand{\eps}{{\varepsilon}}

\def\qmbox#1{\qquad\mbox{#1}\quad}
\def\tr{\mathop{\rm Tr}\nolimits}

\section{ 
\rouge{
$RTT$ presentation for $osp(M|2n)$ Yangians 
}}

Let us consider an $M+2n$ dimensional $\ZZ_2$-graded vector space,
with the $M$ first  
indices bosonic and the $2n$ last ones fermionic.

\rouge{
Define
{ 
\begin{displaymath}
  \label{eq:RPK}
  \displaystyle R(u) = \II + \frac{P}{u} - \frac{Q}{u+\kappa}\;.
\end{displaymath}
}
}
\bleu{
$P$  being the super permutation, and 
$Q=P^{t_1}=P^{t_2}$ being $P$ 
partially transposed.} 
\\
\noindent
The $R$-matrix $R(u)$ satisfies the super Yang--Baxter equation
{ 
\bleu{
\begin{displaymath}
  \label{eq:YBE}
  R_{12}(u) \, R_{13}(u+v) \, R_{23}(v) = R_{23}(v) \, R_{13}(u+v)
  \, R_{12}(u)
\end{displaymath}
}
}
if $2\kappa = (M-2n-2)$, with a 
\rouge{
\emph{graded tensor product}
}.
\\

One defines the Yangian of $osp(M|2n)$ by the generators 
\dgreen{
$$T(u) = \sum_{n \in \ZZ_{\ge 0}}  T_{(n)} \, u^{-n} \qquad T(0)=\II$$ 
}
and the relations
\begin{eqnarray*}
  && R_{12}(u-v) \, T_1(u) \, T_2(v) = T_2(v) \, T_1(u) \, R_{12}(u-v)
  \label{eq:RLL}
  \\
  && T^t(u-\kappa) \, T(u) = \II
\end{eqnarray*}
(i.e. RTT=TTR relations and ``orthogonality relation'') \cite{soya}.

\section{ 
\rouge{ 
  Closed chain integrability
}}

The closed chain monodromy matrix is defined by
\bleu{
\begin{eqnarray*}
  T_{a}(u) = R_{aL}(u) R_{a,L-1}(u) \cdots R_{a
    2}(u) R_{a1}(u)
\end{eqnarray*}
}
Using the Yang--Baxter equation, one proves that the closed chain 
transfer matrices, given by the super trace
$t(u) = \tr_a T_a(u)$, commute for all values of the 
spectral parameter $u$:
\bleu{
{ 
\begin{displaymath}
  \Big[t(u), t(v)\Big] = 0 , \qquad \qquad \forall u,v
\end{displaymath}
}}

\section{ 
\rouge{ 
   Reflection equation and open chain integrability
}}
We consider $K^-(u)\in End(\CC^{M+2n})$, solution of the reflection equation:
\bleu{
\begin{eqnarray*}
  &&
  R_{ab}(u_{a}-u_{b}) \ 
  K^-_{a}(u_{a}) \ 
  R_{ba}(u_{a}+u_{b})\ 
  K^-_{b}(u_{b}) \ = 
  \\
  &&\qquad\qquad
  K^-_{b}(u_{b})\
  R_{ab}(u_{a}+u_{b})\ 
  K^-_{a}(u_{a})\
  R_{ba}(u_{a}-u_{b})
  \label{re}
\end{eqnarray*}
}
Let
\bleu{
\begin{eqnarray*}
  T_{a}(u) = R_{aL}(u) R_{a,L-1}(u) \cdots R_{a 2}(u) R_{a1}(u)
\end{eqnarray*}
}
and
\bleu{
\begin{eqnarray*}
  \hat T_{a}(u) =
  R_{1a}(u) R_{2a}(u) \cdots R_{L-1,a}(u)
  R_{La}(u) 
\end{eqnarray*}
}
The open chain monodromy matrix is the super trace
\bleu{
\begin{displaymath}
  t(u) = \tr_{a} K_{a}^{+}(u)\ T_{a}(u)\
  K^{-}_{a}(u)\ \hat T_{ a}(u)\,, \label{transfer1}
\end{displaymath} 
}
where ${K^+}^t(-\lambda -i\kappa)$ is another solution of the
reflection equation.
Again, as was first proved by Cherednik and Sklyanin using the
Yang--Baxter and reflection equations,
\cite{cherednik,sklyanin},  
$[t(u), t(v)] = 0 , \quad \forall u,v$.

\section{ 
\rouge{ 
  Solutions of the reflection equation
}}

\def\diag{\mathop{\rm diag}\nolimits}

\subsection{
\ddcyan{
Diagonal solutions}}

We solve the reflection equation for $K$ of the form
\ddcyan{
\begin{displaymath}
  \label{eq:diag}
  K(u) = \diag\Big(
  k_1(u),\cdots,k_M(u) \,;\,  k_{M+1}(u),\cdots,k_{M+n}(u) \Big)
\end{displaymath}
}

There are three families of generic diagonal solutions and two
particular cases
\begin{enumerate}
\item[\textbf{D1:}]  
Solutions of $sl(M+2n)$ type, with one free parameter, for
  $M$ even
  \bleu{
  \begin{eqnarray*}
      \label{eq:D1}
      && k_i(u) = 1 \;,
      \\
      && k_{\bar \imath}(u) = \frac{1+cu}{1-cu}\;,
      \qquad \forall\, i\in \{1,...,\frac{M}{2};M+1,...,M+{n}\}
    \end{eqnarray*}
  }
    This solution has no extension to odd $M$. \\
  \item[\textbf{D2:}] Solutions with three different values of $k_l(u)$,
    depending on one free parameter
    \bleu{
    \begin{eqnarray*}
      \label{eq:D2}
      &&
      k_{1}(u) = \frac{1+c_1 u}{1-c_1 u} \;,
      \qquad
      k_{M}(u) = \frac{1+c_{M}u}{1-c_{M}u} \;,
      \qquad
      k_j(u) = 1    \quad \forall\, j \neq 1,M
    \end{eqnarray*}
  }
      where $ (\kappa-1) c_1 c_M + c_1 + c_M = 0$.
    This solution does not hold for $M=0,1$. \\

 \item[\textbf{D3:}] Solutions without any free continuous parameter, but with
  two integer parameters $m_1$, $n_1$, and $c=\frac{2}{\kappa-(2m_1-2n_1-1)}$
  \begin{eqnarray*}
      \label{eq:D3}
      &&
      k_i(u) = k_{\bar\imath}(u) = 1 \qquad \forall\,
      i\in\{1,...,m_1;M+1,...,M+n_1\}
      \nonumber\\[2mm]
      &&
      k_{i}(u) = k_{\bar\imath}(u) = \frac{1+cu}{1-cu} \qquad 
      \qmbox{otherwise}
    \end{eqnarray*}
  \item[\textbf{D4:}] 
    In the particular case of $so(4)$, the solution takes the more 
    general form:
    \bleu{
      \begin{displaymath}
        \label{eq:D4}
        K(u) = \diag\bigg(1\;,\;\frac{1+c_2 u}{1-c_2 u}
        \;,\;\frac{1+c_3 u}{1-c_3 u}\;,\;
        \frac{1+c_2 u}{1-c_2 u}\;\frac{1+c_3 u}{1-c_3 u}
        \bigg)
      \end{displaymath}
    }
    This solution contains the three generic solutions \\
    D1 ($c_2c_3=0$),
    D2 ($c_2+c_3=0$) and D3 ($c_2=c_3=\infty$).
  \item[\textbf{D5:}] In the particular case of $so(2)$, any function-valued
    diagonal matrix is solution.
  \end{enumerate}

\subsection{\dcyan{Antidiagonal and mixed solutions}}
The classification of such solutions is best shown by a few examples.
\\
One finds the two following solutions for $osp(4|2)$ :\\
\textbf{$so$ diagonal :}
\bleu{
\begin{eqnarray*}
&&  \left(
     \begin{array}{cccc|cc}
      {1} & & & & & \\
      & {1} & & & & \\
      & &  {-1} & & & \\
      & & & -{1} & & \\
      \hline
      \Big.
      & & & & k_{5} & \ell_5 \\[1mm]
      & & & & \ell_6 & - k_{5}
     \end{array}
   \right)
\end{eqnarray*}
where $k_{5}^2 + \ell_{5} \ell_{6} = 1$.
\\
\textbf{$sp$ diagonal :}
\begin{eqnarray*}
  && \left(
     \begin{array}{cccc|cc}
       1 & & & 0 & & \\
       & 0 & \ell_2 & & & \\
       & \ell_2^{-1} & 0 & & & \\
       0 & & & 1 & & \\
       \hline
       & & & & 1 & \\
       & & & & & 1
     \end{array}
   \right)
\end{eqnarray*}
}

\noindent
For $osp(2|4)$ the two solutions take the form\\
\textbf{$so$ diagonal :}
\bleu{
\begin{eqnarray*}
 && \left(
    \begin{array}{cc|cccc}
      1 & & & & & \\
      & -1 & & & & \\
      \hline
      & & k_{3} & & & \ell_{3} \\
      & & & k_{4} & \ell_{4} & \\
      & & & \ell_{5} & -k_{4} & \\
      & & \ell_{6} & & & -k_{3}
    \end{array}
  \right)
\end{eqnarray*}
where $k_{3}^2 + \ell_{3} \ell_{6} = 1$ and
$k_{4}^2 + \ell_{4} \ell_{5} = 1$.
\\
\textbf{$sp$ diagonal :}
\begin{eqnarray*}
&&  \left(
    \begin{array}{cc|cccc}
      0 & \ell_1 & & & & \\[1mm]
      \ell_1^{-1} & 0 & & & & \\
      \hline
      & & 1 & & & \\
      & & & -1 & & \\
      & & & & -1 & \\
      & & & & & 1
    \end{array}
  \right)
\end{eqnarray*}
}

\section{ 
\rouge{ 
   Pseudovacuum and one eigenvalue of the transfer matrix for the open chain
}}

We now choose an appropriate pseudo-vacuum, which is an exact
eigenstate of the transfer matrix $t(u)$ of the open chain; 
it is the state with all
``spins'' up, i.e.
\bleu{
\begin{eqnarray*}
\vert \omega_{+} \rangle = \bigotimes_{i=1}^{L} \vert + \rangle
_{i} ~~~~\mbox{where} ~~~~\vert + \rangle = \left (
\begin{array}{c}
1 \\
0 \\
\vdots \\
0 \\
\end{array}
\right)\,\in\,\CC^{M+2n}\,. \label{pseudo}
\end{eqnarray*}
}
Then 
$\qquad
    t(\lambda) ~\vert ~\omega_{+} \rangle  
    = \Lambda^0(\lambda) ~\vert \omega_{+} \rangle \;,
\qquad $
where
\begin{eqnarray*}
  \Lambda^{0}(\lambda) &=& a(\lambda)^{2L} g_{0}(\lambda)
  +
  b(\lambda)^{2L}\sum_{l=1}^{2n+M-2} (-1)^{[l+1]} g_{l}(\lambda)
  +
  c(\lambda)^{2L} g_{2n+M-1}(\lambda) \label{eigen0}
\end{eqnarray*}
with
\dgreen{
\begin{eqnarray*}
  && a(\lambda) =(\lambda+i)(\lambda+i\kappa), \qquad
  b(\lambda) =\lambda
  (\lambda+i\kappa), \qquad
  c(\lambda) =\lambda (\lambda+i\kappa-i)
  \label{numbers}
\end{eqnarray*}
}
the functions $g(\lambda)$ being written as (in the case $osp(2m+1|2n)$,
with $K^{\pm}=\II$)
\begin{eqnarray*}
g_{l}(\lambda) &=& {\lambda (\lambda+ {i\kappa \over 2} -{i \over
2})(\lambda+i\kappa) \over (\lambda+ {i \kappa \over
2})(\lambda+{i l \over 2})(\lambda+{i (l +1)\over 2})},
\qquad
l=0,\ldots,n-1,
\\[2mm]
g_{l}(\lambda) &=& {\lambda (\lambda+ {i\kappa \over 2} - {i \over
2})(\lambda+i\kappa) \over (\lambda+ {i \kappa \over
2})(\lambda+{i n}-{i l \over 2})(\lambda+ in - {i (l+1)\over 2})},
\\ && \hspace{5cm} 
l=n,\ldots,n+m-1
\\[2mm]
g_{n+m}(\lambda) &=&
\frac{\lambda(\lambda+i\kappa)}{(\lambda+i\frac{n-m}{2})
(\lambda+i\frac{n-m+1}{2})}
\qquad
\mbox{if } M=2m+1
\\[2mm]
g_{l}(\lambda) &=& g_{2n+M-l-1}(-\lambda -i\kappa),
\qquad
l=0,1,...,M+2n \label{g2}
\end{eqnarray*}
\rouge{
The dressing consists in  the  insertion of factors $A_l(\lambda)$ to
get the other eigenvalues $\Lambda(\lambda)$ of the transfer matrix
}
\bleu{
\begin{eqnarray*}
  \Lambda(\lambda) &=& a(\lambda)^{2L} g_{0}(\lambda)A_{0}(\lambda)
  + b(\lambda)^{2L}\sum_{l=1}^{2n+M-2}
  (-1)^{[l+1]}g_{l}(\lambda)A_{l}(\lambda)
  \\[2mm] &+& c(\lambda)^{2L}
  g_{2n+M-1}(\lambda)A_{2n+M-1}(\lambda)
\label{eigen}
\end{eqnarray*}
}

The factors $A_l$ take the form
\bleu{
\begin{eqnarray*}
A_{0}(\lambda) &=& \prod_{j=1}^{M^{(1)}}{\lambda+
\lambda_{j}^{(1)}-{i\over 2}\over \lambda+ \lambda_{j}^{(1)}
+{i\over 2}}\ {\lambda-\lambda_{j}^{(1)}-{i\over 2} \over
\lambda-\lambda_{j}^{(1)} +{i\over 2}} \,,\\[2mm]
A_{l}(\lambda) &=& \prod_{j=1}^{M^{(l)}}
{\lambda+\lambda_{j}^{(l)}+{il\over 2}+i \over \lambda+
\lambda_{j}^{(l)} +{il\over2}} \;
{\lambda-\lambda_{j}^{(l)}+{il\over 2}+i\over \lambda-
\lambda_{j}^{(l)} +{il\over 2}} \\[2mm]
& \times & \prod_{j=1}^{M^{(l+1)}}{\lambda+
\lambda_{j}^{(l+1)}+{il\over 2}-{i\over 2}\over \lambda+
\lambda_{j}^{(l+1)} +{il\over 2} +{i\over 2}}\
{\lambda-\lambda_{j}^{(l+1)}+{il \over 2}-{i\over 2} \over
\lambda-\lambda_{j}^{(l+1)} + {il\over 2}+{i\over 2}} \\[2mm] 
 &&l=1,\ldots , n-1 
\end{eqnarray*}
}


Analyticity around the poles introduced in the factors $A_l$ now
imposes the so-called Bethe equations in the $\lambda_i$ :

\bleu{
\begin{eqnarray*}
&&
e_{1}(\lambda_{i}^{(1)})^{2L} = \prod_{\epsilon=\pm1} \prod_{j=1,j \ne
i}^{M^{(1)}} e_{2}(\lambda_{i}^{(1)} -\epsilon \lambda_{j}^{(1)})\
\prod_{
j=1}^{M^{(2)}}
e_{-1}(\lambda_{i}^{(1)} - \epsilon \lambda_{j}^{(2)})\ 
\\[2mm]
&&
1 = \prod_{\epsilon=\pm1} \prod_{j=1,j \ne i}^{M^{(l)}}
e_{2}(\lambda_{i}^{(l)} -\epsilon \lambda_{j}^{(l)})\
\prod_{ \tau = \pm
1}\prod_{ j=1}^{M^{(l+\tau)}}
e_{-1}(\lambda_{i}^{(l)} - \epsilon \lambda_{j}^{(l+\tau)})\ 
\\[2mm]
&& \hspace{2cm}l= 2,\ldots,n+m-1, \;\; l \neq n \\[2mm]
&& 1 = \prod_{\epsilon=\pm1} \prod_{j=1}^{M^{(n+1)}} 
e_{1}(\lambda_{i}^{(n)} - \epsilon \lambda_{j}^{(n+1)})\ 
\prod_{
j=1}^{M^{(n-1)}}e_{-1}(\lambda_{i}^{(n)} - \epsilon \lambda_{j}^{(n-1)})\
\\[2mm]
&&1 = \prod_{\epsilon=\pm1} \prod_{j=1,j \ne i}^{M^{(n+m)}}
e_{1}(\lambda_{i}^{(n+m)} - \epsilon \lambda_{j}^{(n+m)})\
\prod_{j=1}^{M^{(n+m-1)}}
e_{-1}(\lambda_{i}^{(n+m)} - \epsilon \lambda_{j}^{(n+m-1)})\ 
\label{BAE1}
\end{eqnarray*}
}

with 
\dgreen{
$$e_{x}(\lambda)=\frac{\lambda+\frac{ix}{2}}{\lambda-\frac{ix}{2}} \;.$$
}

\section*{Acknowledgments}
This work has been financially supported by the TMR Network
EUCLID: ``Integrable models and applications: from strings to
condensed matter'', contract number HPRN-CT-2002-00325.

\end{document}